\documentstyle[aps,pra,manuscript]{revtex}

\begin{document}
\title{Mimicking a Squeezed Bath Interaction: \\
Quantum Reservoir Engineering with Atoms}
\author{N. L\"utkenhaus \\
J.I. Cirac\\
P. Zoller}
\address{Institut f\"ur Theoretische Physik\\
Universit\"{a}t Innsbruck}
\maketitle

\begin{abstract}
The interaction of an atomic two-level system and a squeezed vacuum leads to
interesting novel effects in atomic dynamics, including line narrowing in
resonance fluorescence and absorption spectra, and a suppressed (enhanced)
decay of the in-phase and out-of phase component of the atomic polarization.
On the experimental side these predictions have so far eluded observation,
essentially due to the difficulty of embedding atoms in a $4 \pi$ squeezed
vacuum. In this paper we show how to ``engineer'' a squeezed-bath-type
interaction for an effective two-level system. In the simplest example, our
two-level atom is represented by the two ground levels of an atom with
angular momentum $J=1/2 \rightarrow J=1/2$ transition (a four level system)
which is driven by (weak) laser fields and coupled to the vacuum reservoir
of radiation modes. Interference between the spontaneous emission channels
in optical pumping leads to a squeezed bath type coupling, and thus to
symmetry breaking of decay on the Bloch sphere. With this system it should
be possible to observe the effects predicted in the context of squeezed bath
- atom interactions. The laser parameters allow one to choose properties of
the squeezed bath interaction, such as the (effective) photon number
expectation number $N$ and the squeezing phase $\phi$. We present results of
a detailed analytical and numerical study.
\end{abstract}
\pacs{42.50.Dv, 32.80Bx, 42.50.Ct, 42.50.Lc}

\narrowtext

\section{Introduction}

The interaction of atomic systems with squeezed light leads to interesting
novel effects in atomic dynamics \cite{parkins93b}. In particular,
Gardiner \cite{gardiner86a} has shown that the atomic Bloch vector of a
two-level system coupled to a squeezed bath, which is characterized by a
mean photon number $N$ and squeezing parameter $M$, obeys the equations 
\begin{eqnarray}
\frac{{\rm d}}{{\rm d}t}\langle S_x\rangle  &=&-\gamma _x\langle S_x\rangle ,
\label{Bloch} \\
\frac{{\rm d}}{{\rm d}t}\langle S_y\rangle  &=&-\gamma _y\langle S_y\rangle ,
\nonumber \\
\frac{{\rm d}}{{\rm d}t}\langle S_z\rangle  &=&-\gamma _z\langle S_z\rangle
-\gamma .  \nonumber
\end{eqnarray}
The decay constants in this equation are given by 
\begin{eqnarray}
\gamma _x &=&\gamma (N+\frac 12-M),  \label{Blochdecayrate} \\
\gamma _x &=&\gamma (N+\frac 12+M),  \nonumber \\
\gamma _z &=&\gamma (2N+1)  \nonumber
\end{eqnarray}
with $\gamma $ is the spontaneous emission rate in free vacuum. We will
refer to Eq. (\ref{Bloch}) as the Gardiner-Bloch Equation. According to Eqs.
(\ref{Bloch},\ref{Blochdecayrate}) the two quadrature components of the
atomic polarization $\langle S_x\rangle $ and $\langle S_y\rangle $ will
decay with different rates $\gamma _x$ and $\gamma _y,$ respectively. In the
limit of large photon number $N\gg 1,$ and maximal squeezing, $M^2=N(N+1),$
the decay of $\langle S_x\rangle $ will be suppressed according to $\gamma
_x\to \gamma /(8N)$ $(\ll \gamma $ ) in comparison to spontaneous emission in
free vacuum. At the same time the decay of $\langle S_y\rangle $ will be
enhanced: $\gamma _y\to 2\gamma N$ ($\gg \gamma )$. As studied in numerous
theoretical papers, these phase sensitive decay rates will also be visible
in the spectrum of resonance fluorescence \cite{carmichael87a,swain96a}, and the
atomic absorption spectrum \cite{ritsch87a}. A study in the context of
the Jaynes-Cummings model has been done in \cite{benaryeh92a}. From a physical
point of view, the suppression of the decay below the free vacuum level is
due to the reduced quantum fluctuations of one of the quadrature components
of squeezed light. Similar effects have been investigated for the case of
three-level systems \cite{ferguson96a,buzek91a}. For experimental realization
one has to include finite bandwidth effects. This has been done for
example in \cite{parkins88a,ritsch88a,parkins90a,parkins90b,yeoman96a}. 

On the experimental side, there have been only few experiments where the
dynamics of atoms in squeezed light has been studied in the laboratory. Most
notable are the experiments by Kimble and coworkers who, for example,
reported experimental observation of the linear intensity dependence \cite
{javanainen90a,geabanacloche89a} of two-photon absorption rate in squeezed
light \cite{georgiades95a}. The predictions of a suppressed and
phase-sensitive decay of the atomic polarization according to Eqs. (\ref
{Bloch},\ref{Blochdecayrate}) have so far eluded observation, essentially
due to the difficulty of embedding atoms in a squeezed vacuum in a complete $%
4\pi $ solid angle. If we denote by $\epsilon $ the fraction of the solid
angle filled by the squeezed vacuum modes then the polarization decay rates
reduce to \cite{gardiner86a} 
\begin{eqnarray*}
\gamma _x &=&\gamma \left[ \epsilon (N+\frac 12-M)+(1-\epsilon )\frac 12%
\right] , \\
\gamma _y &=&\gamma \left[ \epsilon (N+\frac 12+M)+(1-\epsilon )\frac 12%
\right] .
\end{eqnarray*}
Clearly, if $\epsilon $ differs significantly from $1$ the influence of the
squeezed vacuum will be reduced accordingly. One possible solution to
achieve a large effective $\epsilon $ close to $1$ is to consider systems which are
effectively one-dimensional due, for example, to a strong cavity-atom
interactions as proposed (see, e.g. Refs.\cite{parkins93a}  and \cite
{turchette95a}).

In this paper we will show how to {\it ``engineer'' a squeezed-bath type
interaction} leading to a Gardiner-Bloch Equation (\ref{Bloch}) by studying
the dynamics of a{\it \ driven multilevel atom coupled to }$`$`{\it normal
'' vacuum}. In the simplest version we will consider a four-level system:
for example, an angular momentum $J_g=1/2\rightarrow J_e=1/2$ transition
with two (degenerate) ground ($|g_{m=\pm 1/2}\rangle $) and excited states ($%
|e_{m=\pm 1/2}\rangle $), as illustrated in Fig. \ref{sq4level2}. 
If this atomic
transition is driven by $\sigma _{+}$ and $\sigma _{-}$ polarized laser
light, the spontaneously emitted linearly polarized $\pi $-photons emitted
in the transition $|e_{+1/2}\rangle \stackrel{\pi }{\rightarrow }%
|g_{+1/2}\rangle $ and $|e_{-1/2}\rangle \stackrel{\pi }{\rightarrow }%
|g_{-1/2}\rangle $ will {\it interfere} since they are indistinguishable.
For weak driving fields far below saturation we can adiabatically eliminate
the excited states. The dynamics of the two ground states $%
|g_{+1/2}\rangle $, $|g_{-1/2}\rangle $ then obeys a Master equation
with damping terms due to optical pumping processes between the two ground
states. This master equation has a structure analogous to the coupling of a
two-level atoms to a squeezed bath. (\ref{Bloch}) Interfering
processes to mimic squeezed state detection statistics have been used
previously by Wilkens, Lewenstein and
Grochmalicki. \cite{wilkens89a,grochmalicki91a} It is used in the
context of atomic spin measurements in \cite{kuzmich97a}. Reservoir engineering
to influence resonance fluorescence 
by changing the density of states in a cavity has been investigated in
\cite{keitel95a}. In the context of ion motion reservoir engineering
has been shown in \cite{poyatos96a}.

The paper is organized as follows: In the section two we will summarize the
properties of the master equation for a two-level atom coupled to a squeezed
bath for reference and comparison in later sections. In section three we
discuss the reduction of the multilevel master equation to an effective
two-level master equation with squeezed-bath-type couplings of the form (\ref
{Bloch}). Sec. five presents numerical results for resonance fluorescence
and absorption spectra in four level systems, and compares with the
corresponding squeezed bath results. Non-ideal effects, leading to thermal
(as opposed to squeezed) bath coupling, as given for example by the cross
decay terms in spontaneous emission in Fig. \ref{sq4level2}, will be investigated in Sec.
6; in addition we will suggest mechanisms how to suppress these
unwanted effects.

\section{Two level atom in a squeezed vacuum: a summary}

\label{twolevelsystem}

In this section we will briefly review the basic effects and properties of a
two-level system coupled to a squeezed bath. In particular we discuss
solutions of the Gardiner-Bloch Equations \cite{gardiner86a}, and the main
features of the spectrum of resonance fluorescence and the atomic absorption
spectrum. We summarize these results for reference in the following sections.

\subsection{The Gardiner-Bloch Equation}

We consider a two-level atom with ground state $|g\rangle $ and excited
state $|e\rangle $. The coupling of the atom to a squeezed vacuum is
described by the interaction Hamiltonian 
\begin{equation}
H_{{\rm int}}=\sigma \Gamma ^{\dagger }+\sigma ^{\dagger }\Gamma \;.
\end{equation}
with atomic lowering operator is $\sigma =|g\rangle \langle e|$ and a bath
operator $\Gamma ,$ which given in terms of the coupling constants $\kappa
_{k,\lambda }$ and the photon annihilation operator $a_{k,\lambda }$ as 
\begin{equation}
\Gamma (t)=\sum_{k,\lambda }\kappa _{k,\lambda }a_{k,\lambda }e^{-{{\rm {i}}}%
\omega _kt}.
\end{equation}
A broadband squeezed bath centered around the atomic transition frequency $%
\omega _A$ is characterized by the correlation functions 
\begin{eqnarray}
\langle \Gamma ^{\dagger }(t)\Gamma (t^{\prime })\rangle &=&\gamma N\delta
 (t-t^{\prime }),  \label{bdaggerbcorr} \\
\langle \Gamma (t)\Gamma ^{\dagger }(t^{\prime })\rangle &=&\gamma
(N+1)\delta  (t-t^{\prime }),  \nonumber \\
\langle \Gamma (t)\Gamma (t^{\prime })\rangle &=&\gamma Me^{-{{\rm {i}}}%
\phi }e^{-2{{\rm {i}}}\omega _At}\delta (t-t^{\prime }),  \nonumber \\
\langle \Gamma ^{\dagger }(t)\Gamma ^{\dagger }(t^{\prime })\rangle
&=&\gamma Me^{{{\rm {i}}}\phi }e^{2{{\rm {i}}}\omega _At}\delta
(t-t^{\prime }),  \nonumber
\end{eqnarray}
with effective photon number $N$ and the (real) squeezing parameter $M.$
These parameters are restricted by the inequality $M^2\leq N(N+1),$ where the equal sign holds for maximal squeezing. The squeezing phase is
denoted by $\phi $. Here $`$`broadband squeezed bath'' refers to the
assumption that the squeezing bandwidth is larger than the other frequency
scales in the problem (excluding the optical frequency), such as the
spontaneous decay rate $\gamma .$

Knowledge of the correlation functions (\ref{bdaggerbcorr}) allows one to
derive the master equation for the atomic dynamics in the Born-Markov
approximation. In a rotating frame the master equation is \cite
{gardiner86a,carmichael87a} 
\begin{eqnarray}
\frac{{\rm d}}{{\rm d}t}\rho &=&\frac 12\gamma (N+1)\left( 2\sigma \rho
\sigma ^{\dagger }-\sigma ^{\dagger }\sigma \rho -\rho \sigma ^{\dagger
}\sigma \right) \label{masteroriginal}\\
&&+ \frac 12\gamma N\left( 2\sigma ^{\dagger }\rho \sigma -\sigma
\sigma ^{\dagger }\rho -\rho \sigma \sigma ^{\dagger }\right)  \nonumber \\
&&-\gamma Me^{{\rm {i}\phi }}\sigma ^{\dagger }\rho \sigma ^{\dagger
}-\gamma Me^{-{\rm {i}\phi }}\sigma \rho \sigma .  \nonumber
\end{eqnarray}
For non-perfect squeezing we define mean photon numbers $N_1$ and $N_2$
through $N_1(N_1+1)=M^2$ and $N=N_1+N_2$, which allows us to rewrite (\ref
{masteroriginal}) in the form 
\begin{eqnarray}
\frac{{\rm d}}{{\rm d}t}\rho &=&\frac 12\gamma \left( 2\Sigma \rho \Sigma
^{\dagger }-\Sigma ^{\dagger }\sigma \Sigma -\rho \Sigma ^{\dagger }\Sigma
\right)  \label{master2} \\
&&+\frac 12\gamma N_2\left( 2\sigma \rho \sigma ^{\dagger }-\sigma ^{\dagger
}\sigma \rho -\rho \sigma ^{\dagger }\sigma \right)  \nonumber \\
&&+\frac 12\gamma N_2\left( 2\sigma ^{\dagger }\rho \sigma -\sigma \sigma
^{\dagger }\rho -\rho \sigma \sigma ^{\dagger }\right) \nonumber
\end{eqnarray}
with 
\[
\Sigma =\sqrt{N_1+1}\sigma +e^{{\rm i}\phi }\sqrt{N_1}\sigma ^{\dagger }. 
\]
The first line in Eq. (\ref{master2}) are damping terms for ideal squeezing,
while the second and third line in correspond to a thermal reservoir
(background).

In a wave function simulation of the master equation we would interpret $%
\Sigma $, and $\sigma ,$ $\sigma ^{\dagger }$ as quantum jump operators for
the various damping terms in (\ref{master2}). In this quantum jump picture
the master equation is represented by an ensemble of quantum trajectories of
pure system wave functions, where the evolution of states is described by an
effective, non-hermitian, Hamiltonian $H_{{\rm eff}}$, interrupted by
quantum jumps where the wave function undergoes jumps according to the
action of quantum jump operators on the wave function. For a discussion of a
quantum jump picture of two-level atoms coupled to a squeezed bath we refer
to \cite{dum92a}.

With the notation $S_x=\left\langle \sigma ^{\dagger }+\sigma \right\rangle $%
, $S_y=\left\langle \left( \sigma ^{\dagger }-\sigma \right) /i\right\rangle 
$, $S_z=\left\langle P_e -P_g\right\rangle $ for the Blochvector, the Master
 equation (\ref{masteroriginal}) with the phase choice $\phi =0$ is
equivalent to equations (\ref{Bloch}) given in the introduction. The corresponding equation for a
driven system reads 
\begin{eqnarray}
\frac{{\rm d}}{{\rm d}t}\langle S_x\rangle &=&-\gamma (N+\frac 12-M\cos
\phi )\langle S_x\rangle +\gamma M\sin \phi \langle S_y\rangle ,
\label{blochdriving} \\
\frac{{\rm d}}{{\rm d}t}\langle S_y\rangle &=&-\gamma (N+\frac 12+M\cos
\phi )\langle S_y\rangle +\gamma M\sin \phi \langle S_x\rangle +\Omega
_D\langle S_z\rangle ,  \nonumber \\
\frac{{\rm d}}{{\rm d}t}\langle S_z\rangle &=&-\gamma (2N+1)\langle
S_z\rangle -\Omega _D\langle S_y\rangle -\gamma .  \nonumber
\end{eqnarray}
where $\phi = \varphi_S - 2 \varphi_D$ with $\varphi_D$ as phase of
the driving field with Rabi frequency $\Omega_D$ and $\varphi_S$ as
squeezing phase in the same reference frame as $\varphi_D$. The Bloch
equation is given here for the choice $\varphi_D = 0$. 
A consequence of the broken symmetry in the polarization decay is that the
steady state of the driven two-level system becomes dependent on the
relative phase between the squeezing and the driving field.

\subsection{Steady State}

The steady state of the Bloch equation (\ref{blochdriving}) for the driven
two-level system becomes phase dependent if the vacuum is squeezed ($M\neq 0$%
). This means that propagation effects like absorption and dispersion
become dependent on the the relative phase between the driving field
and the squeezing phase. Therefore the steady state might be
interesting to be used as a kind of optical switching where
propagating beams are controlled by the squeezing phase  and the
driving laser. For normal vacuum the steady state is (up to a global
phase) independent of the phase of the driving laser.

\subsection{Spectrum of Resonance Fluorescence}

The spectrum of resonance fluorescence spectrum of a strongly driven
two-level atom is determined by the Fourier-transform $S(\omega )$ of the
stationary two-time correlation function of the atomic dipole: 
\begin{equation}
S_2(\omega )={\rm FT}\left\langle \sigma ^{\dagger }(0)\sigma (\tau
)\right\rangle .  \label{resflucorr}
\end{equation}
Eq. (\ref{resflucorr}) implies that the free field corresponding to
the particular mode for which the
spectrum is observed is not squeezed.

The spectrum of resonance fluorescence for a squeezed bath has been
calculated in Ref. \cite{carmichael87a}. According to the Quantum Regression
Theorem the evolution of the correlation function is governed by the Bloch
equations. Therefore the spectral lines of the resonance fluorescence
spectrum are determined by the eigenvalues of the coefficient matrix in (\ref
{blochdriving}). For strong driving $\Omega _D\gg \gamma _x,\gamma _y$ one
finds a Mollow triplet with phase dependent line-widths and intensities.
(See table \ref{linewidth}.)
 For strong squeezing ($N\to \infty $) the
center line can become arbitrarily close to zero line-width for $\phi
=0$.

\subsection{Atomic Absorption Spectrum}

The absorption spectrum of a weak probe field in presence of a strong
driving field is related to the Fourier Transform of the two-time
correlation function 
\begin{equation}
W(\omega )={\rm FT}\langle [\sigma (\tau ),\sigma ^{\dagger }(0)]\rangle .
\end{equation}
According to Ref. \cite{ritsch87a} $W(\omega )$ shows strong dependence on
the value of $\phi $. For $\phi =0$ it shows a strong peak at the
center and dispersion-like sidebands. The width of the central absorption peak is again dependent on $N$ and narrows to zero
as $N$ grows since its linewidth is the same as in the resonance
fluorescence. Between the center line and the sidebands is a region
with gain. For  $\phi =\pi$ these detuning regions
with gain  are stronger and one finds even for the
center line stimulated emission. The center line is broadened as well
in accordance with the center linewidth of the resonance
fluorescence. 

\section{Derivation of a Squeezed Bath Master Equation for an effective
two-level system}

\label{derivation}

The signature of the squeezed bath coupling in the master equation (\ref
{master2}) is the appearance of a quantum jump operator of the form $\sqrt{%
N_1+1}\sigma +e^{{\rm i}\phi }\sqrt{N_1}\sigma ^{\dagger }$. We interpret
the structure of this operator as arising from interference in the atomic
transition from the lower to the upper atomic level ($\sigma ^{\dagger }$),
and the transition from the upper to the lower state ($\sigma $). An
analogous interference will occur in optical pumping processes between the
two ground states in a four-level system in the emission of $\pi $-polarized
photons.

Let us consider the four-level atom according to Fig.
\ref{sq4level2}.
The two ground states $|g_{m=\pm 1/2}\rangle $ correspond to the two levels $%
|g\rangle ,$ $|e\rangle $ of the previous section. The two upper levels $%
|e_{m=\pm 1/2}\rangle $ will be used as auxiliary levels. The upper and lower
levels are connected by spontaneous decay due to the interaction with the
normal vacuum reservoir. The two decay channels are assumed to give the same
decay rate $\Gamma $ and to give rise to emitted photons of the same
polarization and frequency, so that the two spontaneous emission processes
are indistinguishable. We add two weak resonant laser fields to connect
coherently the levels $|g_{-1/2}\rangle $ to $|e_{+1/2}\rangle $ with a Rabi
frequency $\epsilon _{-}\Omega $ $\left( \ll \Gamma \right) $ and $%
|g_{+1/2}\rangle $ to $|e_{-1/2}\rangle $ with a Rabi frequency $\epsilon
_{+}\Omega $ $\left( \ll \Gamma \right) $. ($\epsilon_+^2 +
\epsilon_-^2 = 1$) Thus, the two ground states are
connected by optical pumping. A first process starts at level $%
|g_{-1/2}\rangle $, transfers via the weak laser field with Rabi frequency $%
\epsilon _{-}\Omega $ to level $|e_{+1/2}\rangle $ from where it decays to
level $|g_{+1/2}\rangle $. The second process, will connect $%
|g_{+1/2}\rangle $ to $|g_{-1/2}\rangle $, due to absorption from the other
laser with Rabi frequency $\epsilon _{+}\Omega $ $\left( \ll \Gamma \right) $%
. As a result we end up with a master equation for the effective two-level
system $|g_{+1/2}\rangle ,$ $|g_{-1/2}\rangle $where the damping terms have
the structure analogous to a squeezed bath coupling. The cross decay terms
corresponding emission of a $\sigma _{+}$ or $\sigma _{-}$ photon will not
interfere, and thus give rise a phase-insensitive background (deviation from
the ideal squeezed-bath type couplings).

\subsection{Adiabatic Elimination in the Four-Level System Master Equation}

We will now derive the master equation for the effective two-level system
starting with the master equation for the four-level system, including the
cross decay (Fig. \ref{sq4level2})  It is given by 
\begin{eqnarray}
\frac{{\rm d}}{{\rm d}t}\rho &=&-{i}{\ \left( H_{\mathrm eff}\rho -\rho
H_{\mathrm eff}^{\dagger }\right) }  \label{master} \\
&&{\ +g_l^2\Gamma \left( \sigma _1+\sigma _2\right) \rho \left( \sigma
_1^{\dagger }+\sigma _2^{\dagger }\right) } \nonumber \\
&&{\ +g_c^2\Gamma \sigma _{-}\rho \sigma _{-}^{\dagger }+g_c^2\Gamma \sigma
_{+}\rho \sigma _{+}^{\dagger }\;} \nonumber 
\end{eqnarray}
with $H_{{\rm eff}}$ an effective Hamiltonian 
\begin{equation}
H_{{\rm eff}}=-{\rm {i}\frac \Gamma 2P_e+\epsilon _{-}\frac \Omega 2\left(
\sigma _{-}^{\dagger }+\sigma _{-}\right) +\epsilon _{+}\frac \Omega 2\left(
e^{-{i}\phi_L }\sigma _{+}^{\dagger }+e^{{i}\phi_L }\sigma _{+}\right) }
\end{equation}
where $\phi_L$ is the relative phase between the two lasers.
The atomic lowering operators are 
\begin{eqnarray*}
\sigma _1 &=&|g_{-1/2}\rangle \langle e_{-1/2}|,\quad \sigma
_2=|g_{+1/2}\rangle \langle e_{+1/2}|, \\
\sigma _{+} &=&|g_{+1/2}\rangle \langle e_{-1/2}|,\quad \sigma
_{-}=|g_{-1/2}\rangle \langle e_{+1/2}| \; .
\end{eqnarray*}
Furthermore, we define a projection operator for the upper states as 
\[
P_e=|e_{-1/2}\rangle \langle e_{-1/2}|+|e_{+1/2}\rangle \langle e_{+1/2}|. 
\]
The Clebsch-Gordan coefficients for the coupling of the upper to the lower
levels are denoted by $g_l$ and $g_c$, respectively. (Compare Fig. \ref{sq4level2}.)

In a parameter regime satisfying $\epsilon _{\pm }\Omega \ll \Gamma $ we can
adiabatically eliminate the upper levels $|e_{-1/2}\rangle $ and $%
|e_{+1/2}\rangle $. \cite{marte93a} For this we introduce the operator 
\[
D=\epsilon _{-}\sigma _{-}+\epsilon _{+}e^{{{\rm {i}}}\phi_L }\sigma _{+}, 
\]
and use the projections onto the upper states, $P_e$, and the lower states, $%
P_g$, to represent the master equation in terms of the evolution of the
upper states, the lower states and the coherence between upper and lower
states. The corresponding equations are 
\begin{eqnarray}
\frac{{\rm d}}{{\rm d}t}P_e\rho P_e=\frac{{\rm d}}{{\rm d}t}\rho _{ee}
&=&-\Gamma \rho _{ee}-{\rm {i}}\frac \Omega 2\left( D^{\dagger }\rho
_{ge}-\rho _{eg}D\right) \\
\frac{{\rm d}}{{\rm d}t}P_g\rho P_e=\frac{{\rm d}}{{\rm d}t}\rho _{ge} &=&-%
\frac \Gamma 2\rho _{ge}+{\rm {i}}\frac \Omega 2\left( \rho _{gg}D-D\rho
_{ee}\right) \\
\frac{{\rm d}}{{\rm d}t}P_g\rho P_g=\frac{{\rm d}}{{\rm d}t}\rho _{gg} &=&-%
{\rm {i}}\frac \Omega 2\left( D\rho _{eg}-\rho _{ge}D^{\dagger }\right) \\
&&+g_l^2\Gamma \left( \sigma _1+\sigma _2\right) \rho _{ee}\left( \sigma
_1^{\dagger }+\sigma _2^{\dagger }\right) +g_c^2\Gamma \sigma _{-}\rho
_{ee}\sigma _{-}^{\dagger }+g_c^2\Gamma \sigma _{+}\rho _{ee}\sigma
_{+}^{\dagger }\;.  \nonumber
\end{eqnarray}
In the adiabatic approximation, $\rho _{ee}$ follows adiabatically the
changes in the coherences described by $\rho _{eg}$ which in turn follow
adiabatically the dynamics of the ground states $\rho _{gg}$. In this
approximation we find for the density matrix restricted to the lower levels, 
$\rho _{gg}$, a master equation which involves two Lindblad terms, 
\begin{eqnarray}
\label{masterfourlevelterm}
\frac{{\rm d}}{{\rm d}t}\rho _{gg}&=&\frac 12 \frac{\Omega ^2}\Gamma
g_l^2 \left( 2 \tilde{\Sigma} \rho \tilde{\Sigma}^\dagger - \tilde{\Sigma}^\dagger \tilde{\Sigma} \rho - \rho \tilde{\Sigma}^\dagger \tilde{\Sigma}
\right)\\
& & + \frac 18 g_c^2 \frac{\Omega ^2}\Gamma \left( 2 \sigma_z \rho
\sigma_z - 2 \rho \right) \;,
\end{eqnarray}
with the jump operator $\tilde{\Sigma}$ given in terms of the Raman operator $\sigma
=|g_{-1/2}\rangle \langle g_{+1/2}|$ as 
\begin{equation}
\tilde{\Sigma} =\left(\epsilon_{+}\sigma  + e^{{\rm {i} \phi }}\epsilon _{-}\sigma ^{\dagger }\right) \; .
\end{equation}
The remaining parts of the four level density matrix are related to $\rho
_{gg}$ by 
\begin{equation}
\rho _{ee}=\frac{\Omega ^2}{\Gamma ^2}D^{\dagger }\rho _{gg}D
\end{equation}
and 
\begin{equation}
\rho _{eg}={\rm {i}\frac \Omega \Gamma D^{\dagger }\rho _{gg}.}
\end{equation}
We make the replacements of table \ref{comparison} to write this master
equation as 
\begin{eqnarray}
\frac{{\rm d}}{{\rm d}t}\rho &=&-{\rm i}\left( H_{{\rm eff}}\rho -\rho H_{%
{\rm eff}}^{\dagger }\right) +  \label{fullmaster} \\
&&+g_l^2\gamma \left( \sqrt{N+1}\sigma +e^{{\rm i}\phi }\sqrt{N}\sigma
^{\dagger }\right) \rho \left( \sqrt{N+1}\sigma ^{\dagger }+e^{-{\rm i}%
\phi }\sqrt{N}\sigma \right)  \nonumber \\
&&+g_c^2\gamma \left( \frac N2+\frac 14\right) \sigma _z\rho \sigma _z\;. 
\nonumber
\end{eqnarray}
with the effective Hamiltonian 
\begin{equation}
H_{{\rm eff}}=-{\rm {i}}\frac \gamma 2\left( g_c^2\left( \frac N2+\frac 14%
\right) +g_l^2\left( N\openone +P_{+}\right) \right) \;.
\end{equation}
The additional jump operator damping term involving 
\[
\sigma _z=P_{+}-P_{-} 
\]
is characteristic for a process which destroys the decoherence between the
two ground state levels. \cite{carmichael93a} Now we also consider a
laser connecting the Raman transition $|g_{-1/2} \rangle \to |g_{+1/2}
\rangle$. We will assume that these lasers are far from any atomic
resonance so that spontaneous emission can be
neglected. (Alternatively, one can consider an AC magnetic field.) The
the Bloch equation is given by
\begin{eqnarray}
\frac{{\rm d}}{{\rm d}t}\langle S_x\rangle &=&-\gamma (N+\frac 12-g_l^2M\cos
\phi_{L} )\langle S_x\rangle +\gamma g_l^2M\sin \phi_{L} \langle S_y\rangle ,
\label{blochfull} \\
\frac{{\rm d}}{{\rm d}t}\langle S_y\rangle &=&-\gamma (N+\frac 12+g_l^2M\cos
\phi_{L} )\langle S_y\rangle +\gamma g_l^2M\sin \phi_{L} \langle S_x\rangle
+\Omega _D\langle S_z\rangle ,  \nonumber \\
 \frac{{\rm d}}{{\rm d}t}\langle S_z\rangle &=&-\gamma g_l^2(2N+1)\langle
S_z\rangle -\Omega _D\langle S_y\rangle -g_l^2\gamma   \nonumber
\end{eqnarray}
where $\Omega_D$ is the Rabi frequency of the Raman transition with
the phase chosen as $\phi_R=0$. 
\subsection{Discussion}
    
\label{mimic}For $g_l=1$ (no cross decay) Eq. (\ref{blochfull}) is
precisely the Bloch equation (\ref{blochdriving}). We can bring the master
equation (\ref{masterfourlevelterm}) into the form (\ref{masteroriginal}) or the Bloch form (\ref{Bloch}) by
making the identifications of table \ref{comparison}.
The two ground levels thus show a dynamics which is analogous to that of a
two-level atom in a squeezed bath. Effectively we $`$`engineer'' with
help of the  upper levels and the laser fields a a
reservoir for
the two ground
levels such that it looks like the coupling to a squeezed bath. (See
figure \ref{environment}.)
This point of view can be
brought out more clearly by writing an interaction Hamiltonian for the lower
states in the adiabatic elimination as 
\begin{equation}
H_{{\rm int}}=\sigma _{gg}\tilde{\Gamma}^{\dagger }+\sigma _{gg}^{\dagger }%
\tilde{\Gamma}
\end{equation}
with $\sigma _{gg}=|g_{1/2}\rangle \langle g_{-1/2}|$ the transition
operator between the two ground states, and the bath operator $\tilde{\Gamma}
$ given in terms of the coupling constants $\kappa _{k,\lambda }$ and the
photon annihilation operator $a_{k,\lambda }$ as 
\begin{equation}
\tilde{\Gamma}=\sum_{k,\lambda }\left[ \left( {\rm {i}}\frac \Omega \Gamma
\right) \left( \kappa _{k,\lambda }\epsilon _{-}a_{k,\lambda }^{\dagger }e^{{%
{\rm {i}}}\omega _kt}e^{-{\rm {i}\omega _Lt}}-\kappa _{k,\lambda
}^{*}\epsilon _{+}e^{-{\rm {i}\phi }}a_{k,\lambda }e^{-{{\rm {i}}}\omega
_kt}e^{{\rm {i}\omega _Lt}}\right) \right] \;.
\end{equation}
The correlation function of this bath operators are the same as the
correlations functions of the operator $\Gamma $ given in (\ref{bdaggerbcorr}%
) of the squeezed bath if one uses the identification of table \ref
{comparison} and takes into account that for the degenerate ground-states
the frequency corresponding to $\omega _A$ vanishes.

In summary, the main elements of the four level system is that the direct
transition between the two levels of the two-level scheme is replaced by a
pumping process involving other atomic levels, and that the driving between
the two ground states is replaced by a Raman coupling.

\section{Discussion and Results: the Ideal Model}

We will now present some examples of the ideal $(g_l = 1)$ behavior of the four level system
to illustrate that it indeed mimics a two-level atom coupled to a squeezed
bath. The following figures were produced using the full master equation for
the four level system of Fig. \ref{sq4level2}. Analytic formulas presented,
however, use the adiabatic eliminated equations for the ground states to
allow an easier comparison to the two-level system. For the parameters used
in the figures, the adiabatically eliminated and the exact calculations are
in excellent agreement.

\subsection{Steady State Solution}

An example of exact correspondence is the steady state solution for the
driven system as investigated by Carmichael, Lane and Walls. \cite
{carmichael87a} We drive the system by a resonant Raman transition with an
effective Rabi frequency $\Omega_D$. A typical steady state is shown
as solid line in figure \ref{steady}.
The dependence on the phase $\phi$ allows light fields, causing the two
internal transitions and the driving, to interact depending on the relative phase
between them. This leads to effects in the propagation of those fields
through a cloud of atoms with the effective four level system as shown
above. The results for the four-level system show the $\phi$-dependence in
agreement with a corresponding two-level system.

\subsection{Absorption Spectrum}

As our first example we study the absorption spectrum of the four level
system. The idea is, in analogy of the investigation concerning the
two-level system, to drive the effective two-level system strongly coupling
the two ground states in a Raman transition which is tuned on
resonance.  Then the weak field absorption spectrum is measured, where the probe field
is represented by another stimulated Raman transition. The stationary
absorption spectrum is then given by the Fourier transform of a two-time
correlation function 
\begin{equation}
W(\omega )\sim {\rm FT}\langle \sigma _{gg}(\tau )\sigma _{gg}^{\dagger
}(0)-\sigma _{gg}^{\dagger }(0)\sigma _{gg}(\tau )\rangle \;
\end{equation}
which can be calculated using the Quantum Regression Theorem.

The absorption spectrum of the four-level system (see figure
\ref{absplotfull} (a) ) is identical to that of the
two-level system in a squeezed bath and we can see  the phase dependence and the sharp peak at the
center for $\phi =0$ (solid line).
This is readily understood since the process involved
in measuring the absorption spectrum makes use only of the system level
dynamics as described by the master equation (\ref{fullmaster}). This
stands in contrast to the situation of the resonance fluorescence
which differs slightly from that of the two-level scheme, as
will be shown in the next subsection. 

\subsection{Resonance Fluorescence}

For resonance fluorescence we drive the system again by a stimulated Raman
transition and detect the fluorescence coming from the transitions $|e_{\pm
1/2}\rangle $ $\stackrel{\pi }{\rightarrow }$ $|g_{\pm 1/2}\rangle $. This
leads, in the adiabatic elimination, to the spectrum given in terms of
ground state correlations as 
\begin{equation}
S(\omega )={\rm FT}\left\langle \left( \epsilon _{+}e^{{\rm {i}\phi }}\sigma
_{gg}^{\dagger }(0)+\epsilon _{-}\sigma _{gg}(0)\right) \left( \epsilon
_{+}e^{-{\rm {i}\phi }}\sigma _{gg}(\tau )+\epsilon _{-}\sigma
_{gg}^{\dagger }(\tau )\right) \right\rangle \;.
\end{equation}
In contrast, the fluorescence spectrum calculated by \cite{carmichael87a} is
based on the Fourier Transform of the correlation function $S_2(\omega )=%
{\rm FT}\left\langle \sigma ^{\dagger }(0)\sigma (\tau )\right\rangle $ (see
Sec. 2), which is based on the assumption that there is no interference in
the detector between the source field and the squeezed vacuum
modes. In our case the fluorescence spectrum is due to a kind of
atomic quadrature correlations (similar to those appearing in the
squeezing spectrum \cite{collett84a}) and there are three additional
terms with respect to the two-level scheme if we use
the four level scheme. However, this does not change the position and width
of the lines in the Mollow triplet. Therefore the line narrowing and the
phase dependence of the line intensities can be observed in the four-level
system. (See figure \ref{resplotfull} (a).)
The resonance fluorescence of the four-level scheme differs from that
of the two-level which is not surprising since its origin of the
can be explained only in the full four-level
scheme and not in the effective two-level scheme of the
groundstates. In this point the resonance fluorescence differs from
the absorption spectrum and the steady state which can be explained
just in terms of the Master equation for the two ground states. 

 We continue to speak of the Mollow
triplet for the four-level system in this parameter regime since only these
three lines have non-negligible intensity. As one leaves the regime of
validity of the adiabatic elimination, where the effective two-level system
and the four-level system are in perfect agreement, one finds, of course, a
richer line structure. (See subsection \ref{validityelim}.)

\section{Results and Discussion: Non-ideal effects}

\label{nonideal} In this section we will address separately two problems
connected to the validity of the effective two-level system master equation
with squeezed-bath type couplings. One is the question of the influence of
dephasing terms in the master equation (\ref{fullmaster}) due to cross
decay. The other is the question of behavior of the system at the onset of
saturation.

\subsection{Effects of Cross Decay}

For a four-level system $J_g=\frac 12\rightarrow J_e=\frac 12$ the
Clebsch-Gordan coefficients are $g_l^2=\frac 13$ and $g_c^2=\frac 23$, so
that most of the spontaneous processes go into the (unwanted) dephasing
part. This ratio improves by using Zeeman sub-levels of higher angular
momentum as upper levels where with help of the AC Stark effect only the
levels with $m=\pm \frac 12$ take part in the dynamics while the other
levels are shifted off resonance by a laser field. For $J_e=\frac 32$ we
then find the slightly more favorable numbers of $g_l^2=\frac 23$ and $g_c^2=%
\frac 13$.

The effect of a non-vanishing $g_c$ on the spectra and the steady state is
quite different. The absorption spectrum and the resonance fluorescence
spectrum show even for $g_c^2=\frac 23$ a strong dependence on $\phi $. (See
figures \ref{absplotfull} and \ref{resplotfull}.)

The line-width of the resonance fluorescence interpolates linearly as a
function of $g_c^2$ between the values valid for the interaction with a
squeezed vacuum and those for the normal vacuum as shown in table \ref
{linewidthfull}.
The steady state variation with $\phi $, especially that of the component $%
S_x$, is more sensitive to the presence of cross decay. To utilize the $\phi 
$ dependence of the steady state one has to find means to suppress this
cross decay to an extent as large as possible (see figure \ref{steady}.)

\subsection{Suppressing the Effects of Cross Decay}
\label{suppressioncrossdecay}
To suppress the dephasing in favor of the squeezed-bath-type effects one can
use destructive interference  of the cross decay in a configuration with more
atomic levels, similar to the one proposed in \cite{zhu96a}.  This scheme employs upper levels Zeeman levels of angular
momentum $J_e=\frac 12$ {\em and} those of angular momentum $J_a=\frac 32$.
(See figure \ref{sq8level}.)
The weak laser fields of frequency $\omega _L$
are now detuned between the upper levels with $J_e=\frac 12$ and those with $%
J_a=\frac 32$ with detuning $\Delta _e=\omega _L-\omega _{eg}$ and $\Delta
_a=-(\omega _L-\omega _{ag})$, respectively. The spontaneous decay rate and
the Clebsch-Gordan coefficients have subscript indicating the level they are
referring to. We can calculate the rate for processes starting and ending in
the same ground level. This rate depends on the detuning and the dipole
elements $d_e$ and $d_a$, referring to the transition between the selected
ground state and the two intermediate upper levels taking part in the
transition respectively. It is given by 
\begin{equation}
\frac 1\tau =\frac{12c^3\epsilon _0|E|^2}{\hbar \omega _L^3}\left| \frac{{%
g_c^{(a)}}^2\Gamma _a}{\Delta _a-{{\rm {i}}}\frac{\Gamma _a}2}-\frac{{%
g_c^{(e)}}^2\Gamma _e}{\Delta _e-{{\rm {i}}}\frac{\Gamma _e}2}\right| ^2\;
\label{tau}
\end{equation}
where we introduced the field strength $|E|$ corresponding to the Rabi
frequencies $\epsilon _{+/-}\Omega $ respectively. We always can choose the
detuning such that $\frac{\Delta _e}{\Delta _a}=\frac{{g_c^{(e)}}_2\Gamma _e%
}{{g_c^{(a)}}^2\Gamma _a}$. This will reduce the cross decay by an order of
magnitude in $\frac{\Gamma _{e/a}}{\Delta _{e/a}}$. A demonstration of
this effect in a similar scheme has been given by Xia, Ye and Zhu
\cite{xia96a}. Any remaining cross
decay can be described, within the validity of the adiabatic elimination, by
effective Clebsch-Gordan coefficients and an effective decay rate.
\cite{footnote}

\subsection{Validity of the Adiabatic Elimination}
\label{validityelim}
Finally, we investigate the behavior of the four-level system if we increase
the Rabi frequencies of the internal transitions. To explore the range of
validity of the adiabatic elimination and the effects of onset of
saturation, we compare line-width and position of the Mollow triplet for the
adiabatically eliminated situation and the full four level system. In the
latter one we characterize the Mollow triplet as the three narrowest lines
in the resonance fluorescence spectrum. We can find their positions and
line-width as eigenvalues of the 15 by 15 matrix appearing in the Bloch
equation of the four level system. The effect of an increased value of $%
\Omega /\Gamma $ for the parameters used to plot the resonance fluorescence
in figure \ref{resplotfull} is shown in Fig. \ref
{eigenplot1}.
Up to $\Omega /\Gamma \approx 0.2$ the line width and positions of the
two-level Mollow triplet follow the curve predicted by the adiabatically
eliminated theory. For higher values of $\Omega /\Gamma $ the same position
is still predicted by both equations but all three lines become narrower in
the full system than expected by the reduced equations.  Other lines of the full four level system prove to be clearly
distinct in their line-width from the ones of the Mollow triplet. We have
checked that the whole appearance of the resonance fluorescence (that is
including the line amplitudes) shows an increasing deviation from the
adiabatically eliminated equations as $\Omega /\Gamma $ increases beyond the
value $\Omega /\Gamma =0.2$ just as expected from the behavior of the
eigenvalues.

\section{Conclusions}

The central result of this paper is that there is a way to engineer an
environment for a two-level system. The basic idea is the use the regime of
adiabatic elimination to link up coherent transitions from system levels to
additional atomic levels with spontaneous transitions from these additional
levels back so system levels. This leads to jump operators in the master
equation which can be designed to purpose. Especially interesting jump
operators can be achieved if one uses, as in the model presented here, that
there are indistinguishable spontaneous processes leading to linear
superpositions of jump operators. With this engineering one
converts the trivial reservoir of normal vacuum modes into the more
sophisticated one of an effective squeezed vacuum filling the whole solid
angle of $4\pi $.

We have illustrated this procedure for the example of a squeezed bath-type
coupling. Within this example we demonstrated this reservoir  can
be engineered by the use of Zeeman sub-levels. The predicted effects of the
squeezed bath coupling on the steady state, the absorptions spectrum
and, with a slight change, the resonance fluorescence have been
recovered in this model. We have shown that one can select spontaneous
decay channels in the multi-level scheme and suppress
others. Therefore one can enhance the effect of engineered
jump operators in the final master equation and suppress other
processes which would partially destroy the desired effects.
 In a numerical
study we have shown the effects of a non-ideal realization resulting
in decoherence processes which are collision like. This study shows
that the predicted effects is robust enough to allow their observation. The
regime of validity of the adiabatic elimination of the upper levels in our
model has been explored. 

Our work presents an approach which allows the observation of
theoretically predicted  effects within reach of  todays experimental
techniques. It encourages the investigation of other reservoir
couplings  which
could be engineered with the ideas presented in this paper.

\acknowledgements
This work has been supported by the  by the Austrian
Science Foundation and by the EU as part of the TMR-Network
``Microlaser and Cavity QED'' ERB-4061-P1-95-1021.

\appendix

\section*{Transition rate for cross decay}

\label{spontem} The value of the detuning which gives destructive
interference can be found using second order perturbation theory. We
consider the subsystem as shown in figure \ref{sqinhibit}.
The Hamiltonian describing the subsystem is given by 
\begin{equation}
H = H_A + H_B + H_{AB} + H_{AL}
\end{equation}
with the system Hamiltonians of the atom, the bath and the laser mode given
by 
\begin{eqnarray}
H_A & = & \hbar \omega_{e} P_{e_{+1/2}} + \hbar \omega_{a} P_{a_{+1/2}} \\
H_B & = & \sum_{k \lambda} \hbar \omega_k a_{k \lambda}^\dagger a_{k \lambda}
\end{eqnarray}
and the interaction Hamiltonians atom--bath and atom--laser given by 
\begin{eqnarray}
H_{AB} & = & \sum_{k \lambda} \left[ \hbar\kappa_{k \lambda}^{(e)} a_{k
\lambda} | e_{+1/2} \rangle \langle g_{-1/2} | + \hbar \kappa_{k
\lambda}^{(a)} a_{k \lambda} | 2a \rangle \langle g_{-1/2} | + {\rm {h.c.} }%
\right] + \\
& & + \sum_{k \lambda} \left[ \hbar \kappa_{k \lambda }^{(e)} a_{k \lambda}
| e_{-1/2} \rangle \langle g_{+1/2} | + \hbar \kappa_{k \lambda}^{(a)} a_{k
\lambda} | 1a \rangle \langle g_{+1/2} | + {\rm {h.c.} }\right] \\
H_{AL} & = & \hbar \Omega_{(e)} | e_{+1/2} \rangle \langle g_{-1/2} | +
\hbar \Omega_{(a)} | a_{+1/2} \rangle \langle g_{-1/2} | + {\rm {h.c.}}
\end{eqnarray}
Here the $\kappa_{k \lambda}^{(e)}, \kappa_{k \lambda}^{(a)}$ are coupling
constants defined in terms of dipole vectors $d_{(e/a)}$, unit vectors for
the electric field $e_{k \lambda}$ and a quantization volume $V$ by 
\begin{equation}
\kappa_{k \lambda}^{(e/a)} = \sqrt{\frac{\omega_{k \lambda}}{2 \hbar
\epsilon_0 V}} (e_{k \lambda}.d_{(e/a)}) \; .
\end{equation}
The Rabi frequencies due to the interaction with the classically treated
laser fields with electric field of amplitude $|E|$ and unit vector $e_{L}$
are 
\begin{equation}
\Omega_{(e/a)} = \frac{2 |E|}{\hbar} (e_{L}.d_{(e/a)}) \; .
\end{equation}
In second order perturbation theory the transition rate for processes
starting and ending in $| g_{-1/2} \rangle$ under spontaneous emission of a
circular polarized photon is given by 
\begin{eqnarray}
\frac{1}{\tau} &=& \frac{4 |E|^2 \omega_L^3}{3\pi c^3 \hbar^3 \epsilon_0}
\left| \frac{|d_a|^2}{\Delta_a} - \frac{|d_e|^2}{\Delta_e}\right|^2 \\
& = & \frac{12 c^3 \epsilon_0 |E|^2}{\hbar \omega_L^3} \left| \frac{{%
g_c^{(a)}}^2 \Gamma_a}{\Delta_a} - \frac{{g_c^{(e)}}^2\Gamma_e}{\Delta_e}%
\right|^2
\end{eqnarray}
Inclusion of the main terms of higher orders is by the resolvent method \cite
{cohen92a} yields the result 
\begin{equation}
\frac{1}{\tau}=\frac{12 c^3 \epsilon_0 |E|^2}{\hbar \omega_L^3} \left| \frac{%
{g_c^{(a)}}^2 \Gamma_a}{\Delta_a-{{\rm {i}}} \frac{\Gamma_a}{2}} - \frac{{%
g_c^{(e)}}^2\Gamma_e}{\Delta_e-{{\rm {i}}} \frac{\Gamma_e}{2}}\right|^2 \; .
\end{equation}
This leads finally to the expression (\ref{tau}). The same result can be
obtained using adiabatic elimination in the master equation for the full
system as shown in figure \ref{sq8level}.

%\bibliographystyle{/home/teazer/norbert/tex/prsty}
%\bibliography{/home/teazer/norbert/tex/norbert}

\begin{thebibliography}{10}
\bibitem{footnote}
Another way to implement this scheme might employ the
Paschen-Back effect which gives rise to level pairs with identical
spontaneous transmission (in an approximation) which are not connected
by cross-decay like spontaneous emission channels due to selection rules.
\end{thebibliography}

\begin{table}[htb]
\center{\ 
\begin{tabular}{l|l|l}
& $\phi = 0$ & $\phi = \pi$ \\ \hline
central peak & $\gamma_x =$ &$ (\gamma_y + \gamma_z)/2 =$ \\
&  $\; \; \gamma \left(N+ \frac{1}{2} - M \right) $ & $\; \; \gamma \left(N+ 
\frac{1}{2} +M \right) $ \\ 
sidebands & $\gamma_y = $ & $ (\gamma_x + \gamma_z)/2 =$ \\
 &  $\; \; \frac{\gamma}{4} \left( 6N+3 +2 M \right) $ & $\; \;
\frac{\gamma}{4} \left(6N+3 -2 M\right)    $
\end{tabular}
\parbox{14cm}{\caption{\label{linewidth} Line-width in the regime of strong
driving. The notations $\gamma_x, \gamma_y, \gamma_z$ refers to the
notation of  equations (\ref{Blochdecayrate}).}}}
\end{table}

\begin{table}[htb]
\begin{tabular}{p{5cm}|c|l}
\multicolumn{2}{c|}{two-level atom in squeezed vacuum} & four level system
\\ \hline
spontaneous emission rate in normal vacuum & $\gamma$ & $\left(
\epsilon_+^2- \epsilon_-^2 \right) \frac{\Omega^2}{\Gamma}$ \\ 
photon number expectation value of squeezed vacuum & N & $\frac{\epsilon_-^2%
}{\epsilon_+^2- \epsilon_-^2}$ \\ 
squeezing parameter & M & $\frac{\epsilon_-\epsilon_+ }{\epsilon_+^2-
\epsilon_-^2}$ \\ 
phase of squeezing & $e^{{\rm {i} \phi}}$ & $e^{{\rm {i} \phi_L}} $
\end{tabular}
\parbox{14cm}{\caption{\label{comparison}
The parameters of the two-level system in a
squeezed vacuum expressed as a function of the parameters of the
mimicking 4 level system}}
\end{table}

\begin{table}[htb]
\begin{tabular}{l|l|l}
& $\phi = 0$ & $\phi = \pi$ \\ \hline
central peak & $\gamma \left(N+ \frac{1}{2} -{g_l}^2 \sqrt{N(N+1)} \right) $
& $\gamma \left( N+ \frac{1}{2} +{g_l}^2 \sqrt{N(N+1)} \right) $ \\ 
sidebands & $\frac{\gamma}{4} \left(2N + 1 + 2 {g_l}^2 ( 2 N + 1 +\sqrt{N(N+1)} )
\right) $ & $\frac{\gamma}{4} \left(2 N + 1 + 2 {g_l}^2 (  2 N + 1 - \sqrt{N(N+1)})
\right)$
\end{tabular}
\parbox{14cm}{\caption{\label{linewidthfull} Line-width in the regime
of strong driving in dependence of effective photon number $N$ and
Clebsch Gordan Coefficient $g_l^2 = 1 - g_c^2$.}}
\end{table}

\begin{figure}[htb]
%\centerline{\epsfig{figure=sq4level2cdr.EPS,width=8cm}}
\parbox{14cm}{\caption{\label{sq4level2}  Realization of a level scheme involving two
ground states and  two upper states which are Zeeman sublevels. The atomic levels are coupled by  
two right and left circular polarized laser fields  with Rabi
frequencies $ \epsilon_+ \Omega$ and $ \epsilon_- \Omega$. These laser
fields are weak in the sense that $\Gamma \gg \epsilon_\pm \Omega$. We
will consider this scheme in two situations: {\em(A)} In
the {\em ideal situation} there is only spontaneous decay along channels with $\Delta m_j =0$. These decays interfere and give rise to squeezed bath like
effects. {\em (B)} In a {\em realistic model}   each upper
level decays with a total decay rate $\Gamma$ with a branching
probability for the decay channels  determined by the Clebsch-Gordan
coefficients $g_c$ and $g_l$. The processes along the cross lines of
decay do not interference since the give rise to right and left
circular polarized photons respectively. The effect of these
cross-decay channels will give rise to collision-like effects of the
reservoir.  In subsection \ref{suppressioncrossdecay} we show how to
suppress the cross-decay. }}
\end{figure}

\begin{figure}[htb]
%\centerline{\epsfig{figure=environment.eps,width=8cm}}
\parbox{14cm}{\caption{\label{environment} Schematic view of the
relation system to reservoir for the two-level and the four level situation. }}
\end{figure}

\begin{figure}[htb]
%\centerline{\epsfig{figure=steady.eps,width=8cm}}
\parbox{14cm}{\caption{\label{steady} Steady state for $N=2.1$,
$\frac{\gamma}{\Gamma}  =1.9  \times 10^{-5}$,     
$\frac{\Omega_D}{\gamma} =5.1$ and for the ideal case of  $g_l = 1$
(solid line) and three other values with  $g_l = 0.99$ (dashed line), $g_l = 0.95$ (dash-dotted line), $g_l = 0.9$ (dotted line).  }}
\end{figure}

\begin{figure}[htb]
%\centerline{\epsfig{figure=absplot.raw.eps,width=8cm}}
\parbox{14cm}{\caption{ \label{absplotfull}  Absorption spectrum
(in arbitrary units)  for
$N=1$,
$\frac{\gamma}{ \Gamma} =\frac{1}{3} \times 10^{-4}$ and
$\frac{\Omega_D}{ \gamma} = 7.1$ and
for (a) the ideal case ($g_l = 1$) and for three values of $g_l = \{ 0.9 \; , \left(\frac{2}{3}\right)^{1/2} \; ,\left(\frac{1}{3}\right)^{1/2}\}$  (sub-plots (b) to (d)). The solid (dotted)
lines show the spectra for $\phi = 0 $ ($\phi = \pi $).
}}
\end{figure}

\begin{figure}[htb]
%\centerline{\epsfig{figure=resplot.raw.eps,width=8cm}}
\parbox{14cm}{\caption{\label{resplotfull} Resonance fluorescence in
 arbitrary units   for
 $N=0.2$,
$\frac{\gamma}{ \Gamma} =7.1 \times 10^{-5}$ and $\frac{\Omega_D}{
\gamma} = 7.1$ and for (a) the ideal case ($g_l =1$) and for three
values of $g_l = \{ 0.9 \; ,\left(\frac{2}{3}\right)^{1/2} \;
,\left(\frac{1}{3}\right)^{1/2}\}$ (sub-plots (a) to (d)). The solid (dotted) line shows the spectrum for
 $\phi = 0$ ($\phi = \pi$). }}
\end{figure}

\begin{figure}[htb]
%\centerline{\epsfig{figure=sq8levelcdr.EPS,width=8cm}}
\parbox{14cm}{\caption{\label{sq8level}  Extended level scheme which is used to
suppress spontaneous decay  between  levels with $m_J =
-\frac{1}{2}$ and $m_J =+\frac{1}{2}$ (cross decay) due to
destructive interference. }}
\end{figure}

\begin{figure}[htb]
%\centerline{\epsfig{figure=eigenplot1.raw.eps,width=8cm}}
\parbox{14cm}{\caption{\label{eigenplot1}  Study of line positions
and width for the full four level system (solid and dash-dotted lines)
in comparison with the results of adiabatically eliminated equations
(dotted lines). The parameter values are $N=0.2$,
$\frac{\Omega_D}{\gamma} = 7.1$ and $g_l = 1$. Compared are the line-width of the center line (top
left), the line-width of the sideband lines (bottom left), the
positions of the three lines (top right). To demonstrate that  the
Mollow triplet is clearly distinguishable from other lines we show the
line-width of the line which is the narrowest after the triplet. (Bottom right.) From this comparison it is clear that for
$\Omega/\Gamma < 0.2$ the approximation made in the adiabatic
elimination of the upper states is valid. }}
\end{figure}

\begin{figure}[htb]
%\centerline{\epsfig{figure=sqinhibitcdr.EPS,width=8cm}}
\parbox{14cm}{\caption{\label{sqinhibit}  Subsystem relevant to investigate the
destructive interference of spontaneous emission along the pumping
transitions. The total decay rates of the upper levels are $\Gamma_{e}$ and
$\Gamma_{a}$. The Clebsch-Gordan coefficients for the spontaneous
decay emitting circular (linear)  polarized photons are denoted by $g_c^{(e)}$
and $g_c^{(a)}$ ($g_l^{(e)}$ and $g_l^{(a)}$). The laser is described by the
electric field amplitude $|E|$ and shown as solid arrow.}}
\end{figure}

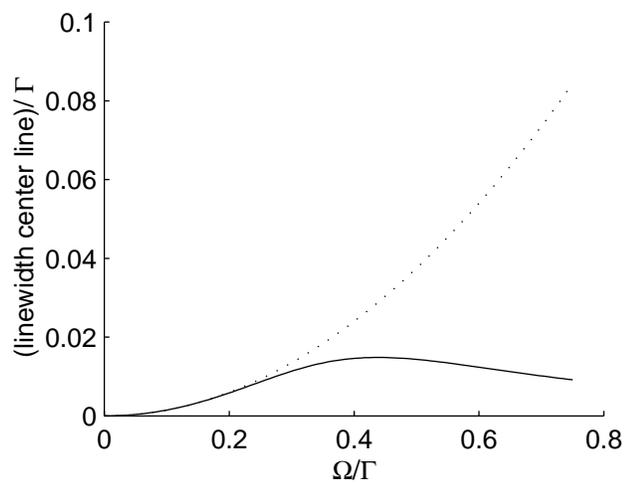
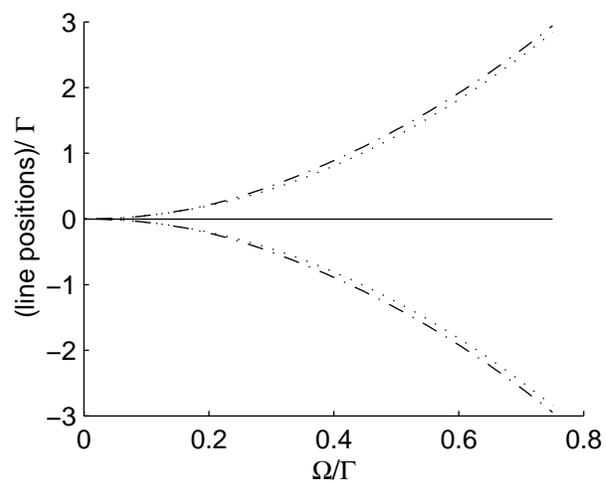
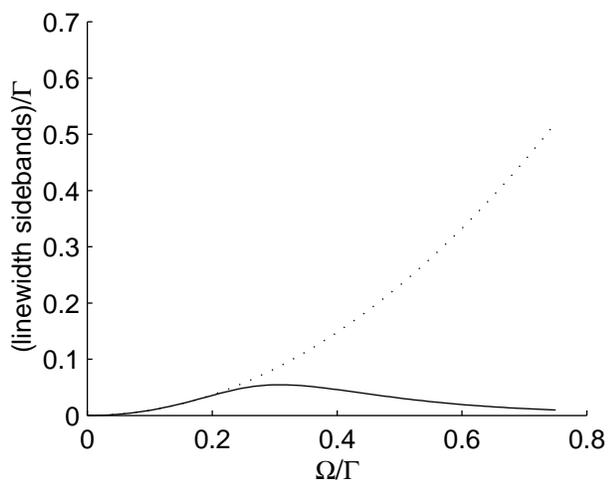
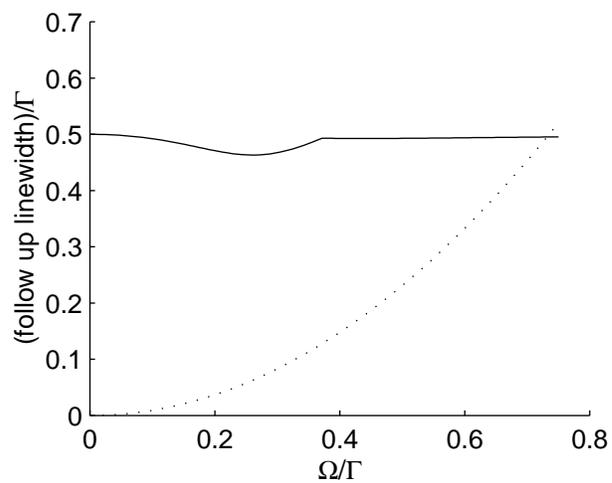
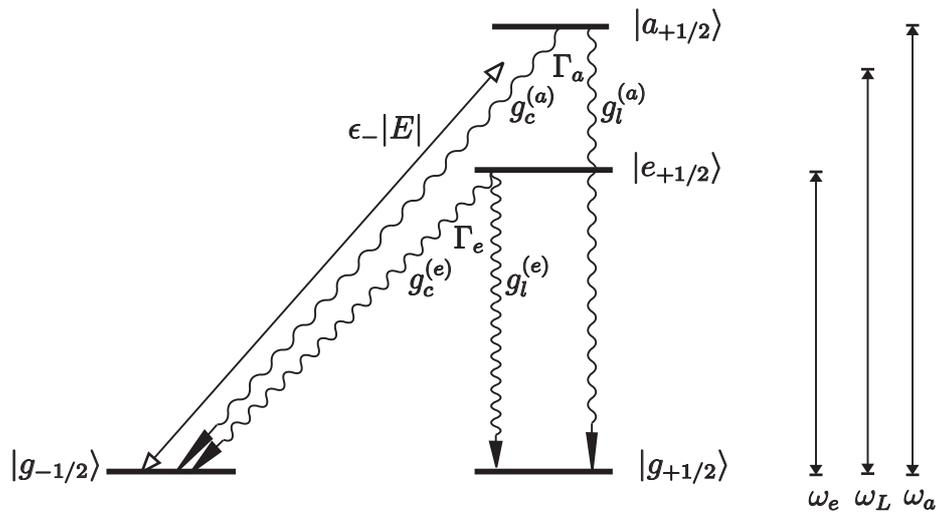
\end{document}